\documentclass[reprint,superscriptaddress]{revtex4-1}
\usepackage{amsmath,amssymb}
\usepackage{graphicx}
\usepackage[sort&compress]{natbib}

\begin{document}

\title{Widefield lensless endoscopy with a multicore fiber}
\author{Viktor Tsvirkun}
\affiliation{Aix Marseille Univ, CNRS, Centrale Marseille, Institut Fresnel UMR 7249, 13013 Marseille, France}
\author{Siddharth Sivankutty}
\affiliation{Aix Marseille Univ, CNRS, Centrale Marseille, Institut Fresnel UMR 7249, 13013 Marseille, France}
\author{G\'{e}raud Bouwmans}
\affiliation{Univ. Lille, UMR 8523 -- Laboratoire de Physique des Lasers Atomes et Mol\'{e}cules, F-59000 Lille, France}
\author{Ori Katz}
\affiliation{Department of Applied Physics, The Selim and Rachel Benin School of Computer Science \& Engineering, The Hebrew University of Jerusalem, Jerusalem 9190401, Israel}
\author{Esben Ravn Andresen}
\email{esben.andresen@ircica.univ-lille1.fr}
\affiliation{Univ. Lille, UMR 8523 -- Laboratoire de Physique des Lasers Atomes et Mol\'{e}cules, F-59000 Lille, France}
\author{Herv\'{e} Rigneault}
\email{herve.rigneault@fresnel.fr}
\affiliation{Aix Marseille Univ, CNRS, Centrale Marseille, Institut Fresnel UMR 7249, 13013 Marseille, France}

\begin{abstract}
We demonstrate pixelation-free real-time widefield endoscopic imaging through an aperiodic multicore fiber (MCF) without any distal opto-mechanical elements or proximal scanners. 
Exploiting the memory effect in MCFs the images in our system are directly obtained without any post-processing using a static wavefront correction obtained from a single calibration procedure.
Our approach allows for video-rate 3D widefield imaging of incoherently illuminated objects with imaging speed not limited by the wavefront shaping device refresh rate. 
\end{abstract}
\maketitle
Fiber-optic microendoscopes allow for minimally invasive high-resolution imaging deep within living organisms.
Over the last decades they continue to gain in versatility with the miniaturization of fiber-based devices and multimodal imaging capabilities.
A new class of such devices, fiber-based lensless endoscopes, operating without any distal optical or mechanical elements, enabled extreme miniaturization of the probe dimensions down to a few hundred micrometers, permitting minimally invasive imaging \cite{Cizmar2012, Choi2012, Andresen2013a, Sivankutty2016b, Papadopoulos2012, Kim2014}.
Image formation in the lensless endoscopes, capable of producing focal planes at various distances from the fiber tip, relies on either raster scanning or widefield modalities. 
In order to reach real-time image acquisition rates, these systems require in the former case ultrafast devices capable of wavefront shaping (typically deformable mirrors) or beam scanning \cite{Andresen2013}, or real-time computation in the latter \cite{Choi2012, Porat2016a}. 

Here we show that, using a slow wavefront-shaping device (spatial light modulator, SLM) and an MCF with weakly coupled cores, it is straightforward to achieve conventional widefield imaging in real time using a single calibration procedure.
Building on the framework developed for imaging through scattering media \cite{Katz2012} and relying on the practically infinite optical memory effect in such MCFs \cite{Sivankutty2016,Stasio2015}, we demonstrate widefield imaging of incoherently illuminated objects.

\begin{figure}[b!]
	\centering
	\includegraphics[width=\linewidth]{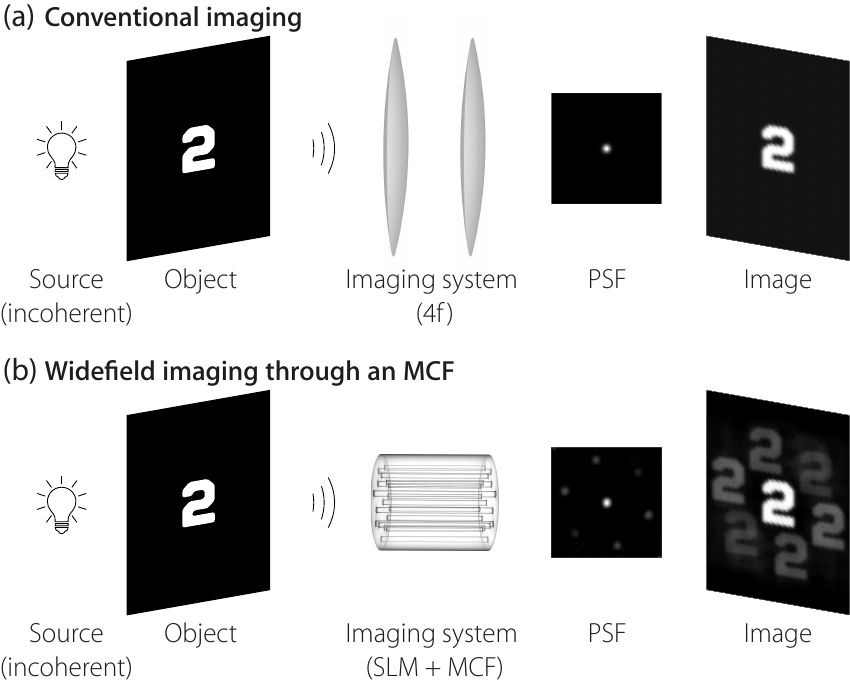}
	\caption{Conceptual view of the experiment for imaging through an MCF bundle. 
		(a) In conventional imaging with lenses, the image is described by a convolution of the object with the PSF of the imaging system.
		(b) In imaging using a combination of the MCF and an SLM, the same principle can be used to describe the image formation.
		A wavefront shaping device (SLM; not shown) is employed to pre-compensate for the inter-core phase dispersion and to correct the wavefront in order to get a focal plane at a desired distance from the MCF tip.
		}
	\label{fig:img1}
\end{figure}

A conceptual illustration for our technique with the corresponding numerical simulation is depicted in Figure \ref{fig:img1}. 
The schematic in Fig.~\ref{fig:img1}a represents a conventional widefield microscope where light, scattered from an incoherently illuminated object, gets collected by a 4\textit{f} system, thus forming an image on the other side of it.
For such an imaging system with a lateral intensity point spread function $\mathrm{PSF}(\vec{r})$, image intensity distribution $I(\vec{r})$ is related to the object $O(\vec{r})$ through a convolution operation: $I(\vec{r}) = O(\vec{r}) \ast \mathrm{PSF}(\vec{r})$.

Typically, an MCF acts as an imaging conduit, transmitting object information from one fiber endface to another. 
In earlier implementations, the individual cores sample the object directly \cite{Gmitro1993, Klimas2014}, giving rise to two important restrictions : i) pixelation due to inter-core separation and ii) imaging restricted to fiber facet itself, putting the probe in contact with the sample. 
The ability to operate at flexible working distances is highly desirable in context of endoscopic applications.
In our previous works \cite{Andresen2013, Andresen2013a}, we reported MCF devices meeting both of the mentioned requirements.
The combination of a wavefront shaping device and an MCF can effectively function in a manner analogous to the common 4\textit{f} system.
A considerable difference being its PSF, which exhibits significant side lobes.
They arise from the fact that the pupil of an MCF is segmented and such discontinuities give rise to prominent side lobes \cite{Sheppard:12}. 
This in turn affects the image, transmitted through the system, as illustrated in Fig.~\ref{fig:img1}b, resulting in ghost images of the object.
Recently we showed \cite{Sivankutty2016} that the randomness in the MCF core positions can lead to a considerable reduction of the side lobes intensity in the PSF of such an imaging system.
A simulation of an imaging experiment with such an aperiodic MCF is shown in  Fig.~\ref{fig:img1}b, where the PSF is calculated given the cores distribution in the real fiber used throughout this work.
While the images exhibit replicas of the object, their intensity is at least 2.5 times lower compared to the central image.
Commercial fiber bundles do exhibit variations in core geometry and spacing in view of decreasing the inter-core coupling \cite{Chen2008},	hence we expect a larger reduction in the side lobe intensity \cite{Stasio2015}. 

Implementations of the MCF with wavefront shaping, reported so far, have all employed raster scanning for imaging \cite{Thompson2011, Andresen2013, Andresen2013a, Stasio2015}.
Unlike in endoscopes based on multimode fibers (MMF), this becomes trivial in MCFs due to their very large memory effect \cite{Sivankutty2016}. 
As there is little or no cross-talk between fiber cores, the transmission matrix of an MCF is practically diagonal. 
Hence, any phase gradient at the proximal end is preserved during light propagation through the fiber. 
This phenomenon has been employed to remotely scan the beam by applying a global tip-tilt on the composite wavefront entering the MCF. 
Since these are relatively simple patterns, the SLM can now be decoupled from having to perform the scanning and conventional galvanometer-based scanners were employed allowing imaging rates of 11 fps \cite{Andresen2013}. 

In this Letter, we propose and demonstrate the extension of the memory effect of MCFs to the widefield imaging framework. 
The reasoning is relatively straightforward, since a translation of a point source in the object plane would result in a phase gradient which is preserved through the fiber (linear in case of the transverse plane and quadratic for the axial plane).  
This results in an accurate mapping of any object shift to the image. 
These concepts are reminiscent of earlier experiments in scattering media \cite{Vellekoop2010, Katz2012}, and the additional key advantage in the case of the MCF being that the memory effect does not play a limiting role in the imaging process.

\begin{figure}[tb!]
	\centering
	\includegraphics[width=\linewidth]{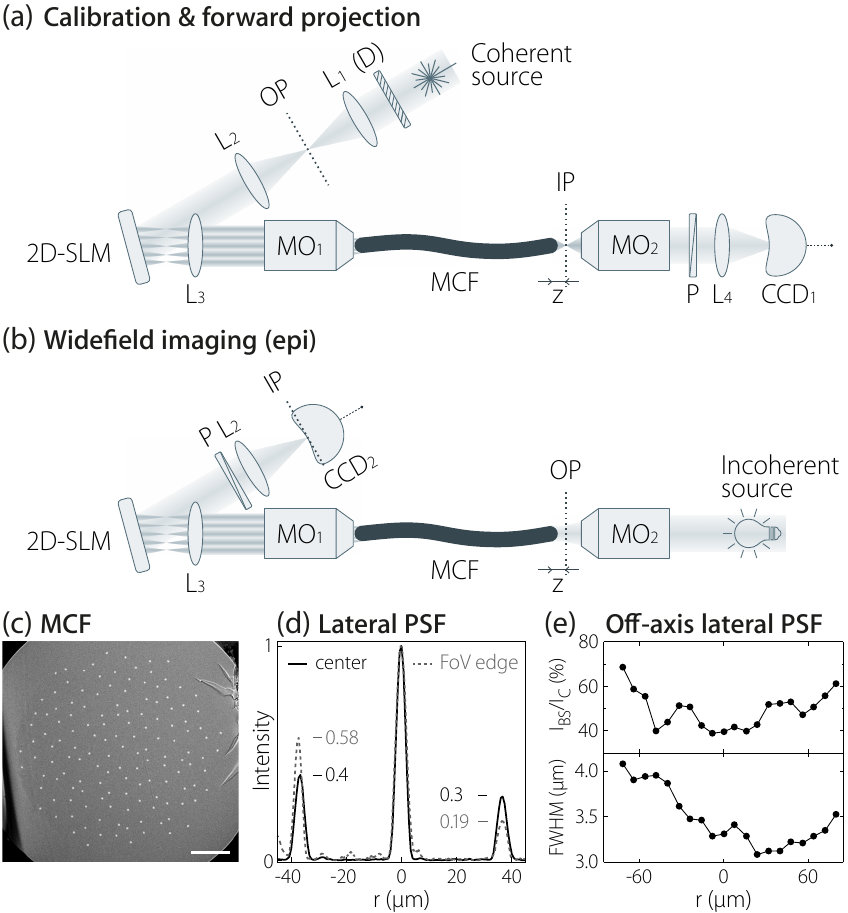}
	\caption{Schematic of the experimental set-up used for system calibration and forward image projection (a), and for the widefield imaging in the epi direction (b).
		(c) SEM image of the aperiodic MCF bundle, used in the experiment (see text for details). 
		Scale bar 50~$\mu$m.
		(d) Measured PSF profile at $z = 600$~$\mu$m from the MCF distal end in the center (solid) and on the edge (dashed) of the FoV.
		(e)	Off-axis PSF evolution over the FoV: ratio between the central peak $I_{\mathrm{C}}$ and the brightest speckle $I_{\mathrm{BS}}$ intensities (top); FWHM of the central peak (bottom).
	}
	\label{fig:img2}
\end{figure}

We focus on the experimental evaluation of such widefield incoherent imaging employing the memory effect to enhance speed and simplicity.
The novel design and the original fabrication approach of this fiber were previously reported in \cite{Sivankutty2016}.
Figure~\ref{fig:img2}c depicts the fiber fabricated with the following parameters: individual core diameter $d_0 = 3.2$ $\mu$m, its numerical aperture $\mathrm{NA} = 0.18$, master triangular lattice pitch $\Lambda = 20$ $\mu$m, and the randomness parameter PR~$\approx 0.22$ (see \cite{Sivankutty2016} for details).
The outer diameter of the probe is about 360~$\mu$m (Fig.~\ref{fig:img2}c) and the length of the fiber used in this experiment was 40~cm, ensuring no considerable inter-core group delay dispersion \cite{Andresen2015}.

A simplified scheme of the experimental set-up, used for the widefield imaging with this MCF, is shown in Figs.~\ref{fig:img2}a,b.
The laser beam from a femtosecond source (Amplitude Syst\`{e}mes t-Pulse, $\lambda = 1030$~nm, 180~fs, 50~MHz) is extended with a telescope (L{\footnotesize 1}, L{\footnotesize 2}) to overfill the aperture of the 2D liquid crystal SLM (Hamamatsu X10468-07).
The latter is used to shape the segmented wavefront entering the MCF via its proximal endface.
During the initial system calibration step, a microlens segment inscribed on the SLM for each individual fiber core produces a focal spot at the focal length $\mathrm{f}_{\mathrm{ml}}$ from the SLM face, and is spatially scanned around its initial position in order to optimize the coupling into the corresponding core.
The array of the optimized focal spots is then imaged onto the MCF proximal endface via a system of lenses (L{\footnotesize 3}, MO{\footnotesize 1}). 
A focal plane IP is created at a distance $z = 600$ $\mu$m away from the MCF distal end and imaged via another telescope system (MO{\footnotesize 2}, L{\footnotesize 4}) onto a camera (CCD{\footnotesize 1}) for calibration and testing purposes. 
Output polarization from different cores of such non polarization maintaining MCF is arbitrary \cite{Sivankutty2016a}, therefore we employ a linear polarizer (P, Thorlabs LPNIR100) to discard any concomitant effects.
After the initial system calibration and compensation of the distal wavefront for the intrinsic MCF phase distortion, one obtains a characteristic PSF (Fig.~\ref{fig:img2}d) comprising a central spot surrounded by six dimmer replicas distributed on a circumference with $r \approx 37$~$\mu$m.
In the linear imaging regime, the brightest speckle contrast relative to the central peak ($I_{\mathrm{BS}}/I_{\mathrm{C}}$) is 0.4, measured in the field of view (FoV) center (Fig.~\ref{fig:img2}d).
We verify the PSF variation across the FoV, in Fig.~\ref{fig:img2}e we summarize such measurements for off-axial points, showing the variation of PSF FWHM less than 1~$\mu$m and $I_{\mathrm{BS}} \approx$ $0.6 I_{\mathrm{C}}$ on the FoV edge, which should not drastically decrease the imaging performance of the widefield technique.

\begin{figure}[tb!]
	\centering
	\includegraphics[width=\linewidth]{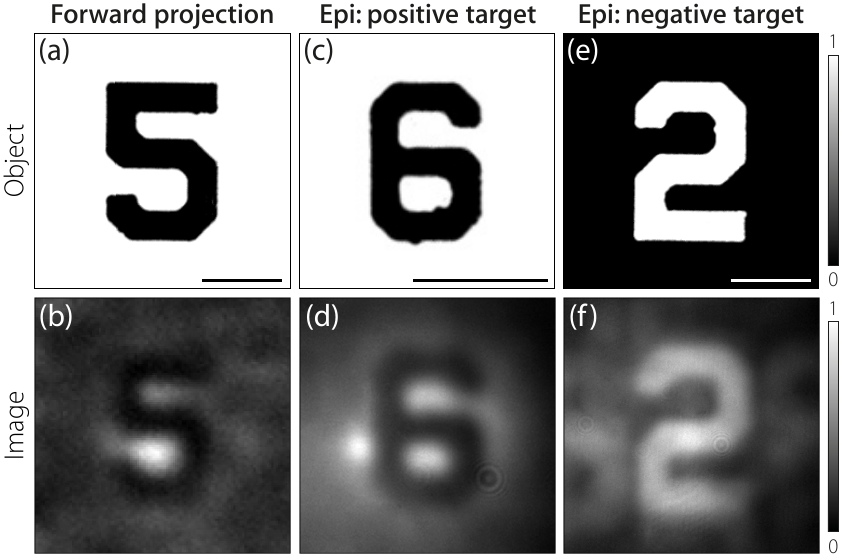}
	\caption{Widefield imaging examples in forward (a,b) and epi (c--f) configurations using USAF target (see text for details).
		(a,b) Forward projection.
		Epi imaging of (c,d) positive and (e,f) negative objects.
		(a,c,e) are optical microscope images of the corresponding target objects; scale bars are 20 $\mu$m.
	}
	\label{fig:img3}
\end{figure}

Using this fiber, we perform a series of proof-of-concept experiments, described in the following.
The experimental set-up, used for the fiber calibration (Fig.~\ref{fig:img2}a), features two conjugate focal planes: OP and IP (where a particular distance $z$ for the IP can be flexibly chosen during the calibration step).
Unlike the calibration step, where it is required to have a spatially and temporally coherent source, the following imaging experiments are performed with spatially incoherent illumination of the object (nevertheless, the bandwidth of the illumination source has to be smaller than the speckle spectral correlation width).
We now perform an incoherent projection of an amplitude mask from the OP to the IP plane; we denote such operation as forward projection.
The related experiment is performed with a reflective object from United States Air Force (USAF) resolution chart (Fig.~\ref{fig:img3}a).
The phase mask on the 2D-SLM is the same used to correct for the intrinsic MCF phase distortion and does not change throughout the experiment, unless one wishes to switch the projection to another working plane (different $z$). 
A number '5' target (object height $\approx 39$~$\mu$m) was placed in the OP and illuminated incoherently by placing a rotating diffuser (D in Fig.~\ref{fig:img2}a) between the object plane and the laser source.
The position of the diffuser along the beam propagation direction is chosen to create in the OP a sufficiently large illumination area, slightly exceeding the dimensions of the used mask. 
The measured projection in the IP is shown in Fig.~\ref{fig:img3}b; it is clear that the set-up performs like a conventional 4\textit{f} imaging system with only a single calibration aided by the memory effect.   
As the SLM does not need to be updated any further, we can easily perform high speed image acquisition (exposure time for the presented example was 30~ms). 
For a real-time imaging experiment of a moving target, see \textbf{Visualization 1}. 

Next, we perform a widefield imaging experiment in the epi direction, using the modified set-up shown in Fig.~\ref{fig:img2}b.
In this configuration the spatially incoherent illumination (the same as for the forward projection experiment) reaches the sample plane (OP) after passing the MO{\footnotesize 2}, and the IP is matched directly to the CCD{\footnotesize 2} camera plane (note that the physical locations of IP and OP are switched as compared to the forward projection experiment).
A linear polarizer (P) is used after the MCF proximal end in the same scope as in the forward projection set-up, as we measure only the scalar transmission matrix \cite{Tripathi2012}.
We use two types of USAF targets -- positive and negative -- to compare the operation of our imaging system in different sample configurations, particularly in terms of the influence of the side lobes.
In the first case we use a positive target consisting of number '6' (Fig.~\ref{fig:img3}c); object height is $\approx 23.5$ $\mu$m, which is almost 3 times less compared to the distance between the side lobes of the system PSF. 
This results in an image on relatively homogeneous background (Fig.~\ref{fig:img3}d) without any overlaying ghost replicas.
In the case of negative target (Fig.~\ref{fig:img3}e), given the object height $\approx 39$~$\mu$m and the distance between replicas, the object convolution with the system PSF results in ghost images of the object that begin to overlap with the central (brightest) '2'.
Unlike in TPEF microscopy \cite{Sivankutty2016} where this weak background is screened by the inherent nonlinearity and does not contribute to image formation, these now result in ghost images albeit with reduced intensity. 
As it can be seen from Fig.~\ref{fig:img3}f, the image associated with the central lobe is still the brightest one and can be easily distinguished from the background generated by the side lobes and any weak speckle. 

We note that earlier results from our group on the same MCF achieved a FoV of 120~$\mu$m (0.064$\pi$), whereas in the present work the illumination area was restricted to cover a FoV of $\approx$ 50~$\mu$m (0.03$\pi$) due to lower experimental SNR.

\begin{figure}[t!]
	\centering
	\includegraphics[width=\linewidth]{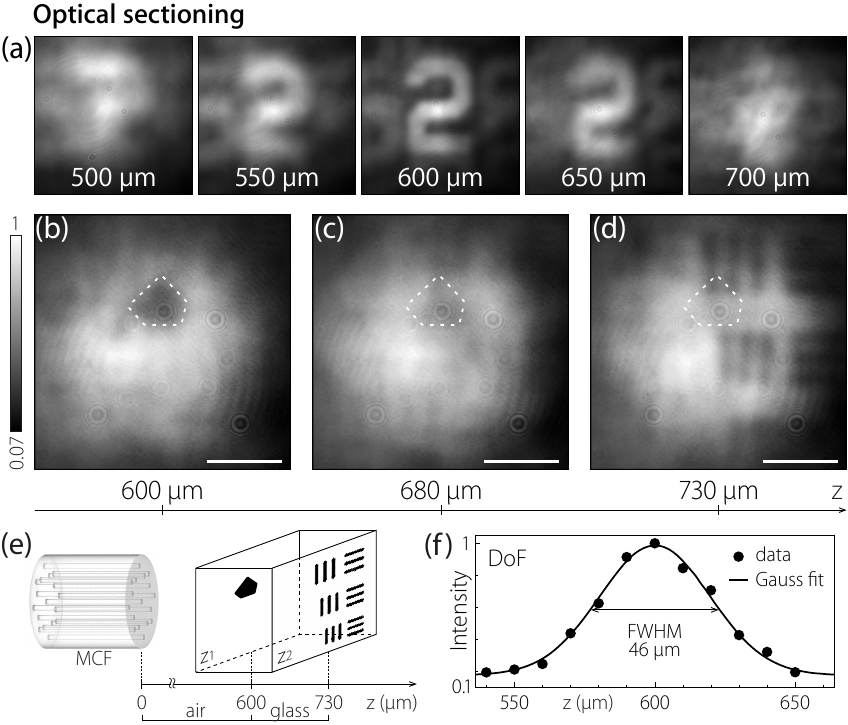}
	\caption{Optical sectioning experiment.
		(a) Images of several distal focal planes (effective focal distance $z$ is indicated on each image) for Fig.~\ref{fig:img3}e object, placed at $z = 600$~$\mu$m.
			(b--d) Images of several distal focal planes within a quasi-3D sample [see (e) and text for details]. Scale bar 20~$\mu$m.
				(d) Schematic of the sample used in the (b--d) experiment.
		(e) Measured depth of field of the imaging system. 
	}
	\label{fig:img4}
\end{figure}

Another advantage of the direct imaging approach with an SLM over speckle-correlation based techniques \cite{Porat2016a} for widefield imaging is that our approach offers a degree of optical sectioning due to its 3D transfer function \cite{streibl1985}.

We performed an experimental measurement of the depth of field (DoF) in the following manner. 
After distortion compensation, the calculated phase pattern is displayed on the SLM in order to obtain a focal plane at $z = 600$~$\mu$m from the MCF distal end. 
Next, a diffraction-limited point source is placed at different $z$ and the DoF is evaluated from the stack of images measured at CCD{\footnotesize 2}, cf.\ Fig.~\ref{fig:img2}b. 
This results in a Gaussian distribution (Fig.~\ref{fig:img4}f) with FWHM~$= 46$~$\mu$m, which is in qualitative agreement with the expected DoF (32~$\mu$m). 
Then, using the same experimental layout as in Fig.~\ref{fig:img2}b and the same test object as in Fig.~\ref{fig:img3}e, placed into the OP at $z = 600$~$\mu$m away from the MCF distal tip, we show in and out of focus images of the target (Fig.~\ref{fig:img4}a) by switching the focal planes through displaying on the SLM the respective differential phase patterns with no mechanical translation of the sample or the fiber.  
As expected, the object appears in focus and then is defocused on the CCD{\footnotesize 2} plane (IP).
For the full stack of 12 imaged focal planes between $z = 500$ and 700~$\mu$m, see \textbf{Visualization 2}.

Considering the measured $\mathrm{DoF}$, we further demonstrate proof of concept optical sectioning experiments (Fig.~\ref{fig:img4}a) with a 3D phantom, schematically depicted in Fig.~\ref{fig:img4}e.
It consists of 50~nm thick layer of gold flakes, deposited onto a 130~$\mu$m thick glass cover slip (\#1). 
From the bottom side of the latter, we stack a positive USAF target (Thorlabs R1DS1P) in a manner that the separation between two metal layers is equal to the cover slip thickness.
Therefore, when introduced into widefield imaging set-up, the gold flakes layer is situated at $z${\footnotesize 1}~$ = 600$~$\mu$m away from the MCF distal end, whereas the USAF target layer is at $z${\footnotesize 2}~$ = 730$~$\mu$m (Fig.~\ref{fig:img4}e).
By adjusting the phase mask on the SLM in order to translate the endoscope focal plane along $z$ axis, we could acquire clear images at several $z$ without moving either the fiber probe or the sample. 
Images of the two planes of interest -- with the gold flake (Fig.~\ref{fig:img4}b) and with USAF target group 7 elements 3--5 (Fig.~\ref{fig:img4}d), plus an intermediate plane at $z = 680$~$\mu$m (Fig.~\ref{fig:img4}c) are shown with the overlaying white dashed lines indicating the outline and the position of the gold flake. 
Remarkably, we are able to focus on both these planes remotely using the SLM, demonstrating pixelation-free and 3D-resolved imaging. 
For the full stack of 8 focal planes, imaged between $z = 600$ and 730~$\mu$m, see \textbf{Visualization 3}. 
We believe this is the first ever demonstration of widefield 3D-resolved imaging at multiple depths in lensless endoscopy employing linear contrast and we expect this to be significantly valuable in the context of imaging 3D structures.
In the current implementation, the presented widefield lensless endoscope is not resilient to fiber bending which would change the phase dispersion within the MCF. 
However, solutions under development may eventually permit real-time compensation of these effects, see e.g. \cite{Andresen2016} and references therein.

We have demonstrated real-time pixelation-free widefield imaging through an aperiodic MCF using the optical memory effect without any distal opto-mechanical elements and requiring no further post-processing. 
Both the forward projection of an amplitude mask from the proximal to distal side of the endoscopic system and epi widefield imaging were shown.
By employing wavefront control at the fiber proximal end, we showed that our system is capable of optical sectioning -- producing clear images of different focal planes within a  quasi-3D sample, and it does so without the need to physically displace either the MCF or the sample.

In the future, we expect that widefield imaging could be combined with nonlinear imaging, e.g. TPEF imaging in a lensless endoscope \cite{Andresen2013}. Fluorescence sources or another broadband illumination whose bandwidth is narrower than the MCF's speckle spectral correlation bandwidth can be also used without significantly affecting
the imaging performance (see \cite{Porat2016a} for details). Considering the robustness, fabrication simplicity of the presented aperiodic MCF with low inter-core coupling, as well as facilitated ultrashort pulse delivery and the small effective probe diameter (an order of magnitude smaller compared to the existing scanning endoscope solutions), such a fiber emerges as a promising candidate in the scope of miniaturized imaging systems. 

\section*{Funding}
Agence Nationale de la Recherche (ANR) (ANR-11-INSB-0006, ANR-10-INSB-04-01,  ANR-14-CE17-0004-01); 
Aix-Marseille Universit\'{e} (ANR-11-IDEX-0001-02); 
Universit\'{e} Lille 1 (ANR-11-LABX-0007, ANR-11-EQPX-0017, CPER P4S R\'{e}gion Nord Pas-de-Calais);
Institut National de la Sant\'{e} et de la Recherche M\'{e}dicale (Inserm) (PC201508); 
SATT Sud-Est GDC Lensless endoscope; 
European research council grant no. 677909;
The Azrieli foundation;
CNRS/Weizmann NaBi European Associated Laboratory.

Authors thank Juan de Torres for providing the gold flakes sample.

\end{document}